\documentclass{ieeeaccess}

\usepackage{bm}

\usepackage{cite}
\usepackage{amsmath,amssymb,amsfonts}
\usepackage{algorithmic}
\usepackage{graphicx}
\usepackage{textcomp}

\def\BibTeX{{\mathrm B\kern-.05em{\sc i\kern-.025em b}\kern-.08em
    T\kern-.1667em\lower.7ex\hbox{E}\kern-.125emX}}

\begin{document}
\history{Date of publication xxxx 00, 0000, date of current version xxxx 00, 0000.}
\doi{10.1109/ACCESS.2019.DOI}

\title{Deep Learning for CSI Feedback Based on Superimposed Coding}
\author{\uppercase{Chaojin Qing}\authorrefmark{1}, \IEEEmembership{Member, IEEE},
\uppercase{Bin Cai}\authorrefmark{1}, \uppercase{Qingyao Yang}\authorrefmark{1}, \uppercase{Jiafan Wang}\authorrefmark{2}, \uppercase {and Chuan Huang}\authorrefmark{3},
\IEEEmembership{Member, IEEE}}
\address[1]{School of Electrical Engineering and Electronic Information, Xihua University, Chengdu, 610039, China. (e-mail: qingchj@ uestc.edu.cn)}
\address[2]{Synopsys Inc., 2025 NE Cornelius Pass Rd, Hillsboro, OR 97124, USA.}
\address[3]{National Key Laboratory of Science and Technology on Communications, University of Electronic Science and Technology of China, Chengdu, 611731, China.}

\tfootnote{This work is supported in part by the Key Projects of Education Department of Sichuan Province (Grant 15ZA0134), the Special Funds of Industry Development of Sichuan Province (Grant zyf-2018-056), the Major Special Funds of Science and Technology of Sichuan Science and Technology Plan Project (Grant 19ZDZX0016), the Key Scientific Research Fund of Xihua University (Grant Z1120941) , and the National Natural Science Foundation (Grant 61501093) of China.}

\markboth
{Chaojin Qing \headeretal: Preparation of Papers for IEEE ACCESS}
{Chaojin Qing \headeretal: Preparation of Papers for IEEE ACCESS}

\corresp{Corresponding author: Chaojin Qing (e-mail: qingchj@ uestc.edu.cn).}

\begin{abstract}
Massive multiple-input multiple-output (MIMO) with frequency division duplex (FDD) mode is a promising approach to increasing system capacity and link robustness for the fifth generation (5G) wireless cellular systems. The premise of these advantages is the accurate downlink channel state information (CSI) fed back from user equipment. However, conventional feedback methods have difficulties in reducing feedback overhead due to significant amount of base station (BS) antennas in massive MIMO systems. Recently, deep learning (DL)-based CSI feedback conquers many difficulties, yet still shows insufficiency to decrease the occupation of uplink bandwidth resources. In this paper, to solve this issue, we combine DL and superimposed coding (SC) for CSI feedback, in which the downlink CSI is spread and then superimposed on uplink user data sequences (UL-US) towards the BS. Then, a multi-task neural network (NN) architecture is proposed at BS to recover the downlink CSI and UL-US by unfolding two iterations of the minimum mean-squared error (MMSE) criterion-based interference reduction. In addition, for a network training, a subnet-by-subnet approach is exploited to facilitate the parameter tuning and expedite the convergence rate. Compared with standalone SC-based CSI scheme, our multi-task NN, trained in a specific signal-to-noise ratio (SNR) and power proportional coefficient (PPC), consistently improves the estimation of downlink CSI with similar or better UL-US detection under SNR and PPC varying.
\end{abstract}


\titlepgskip=-15pt
\maketitle

\section{Introduction}
\label{sec:introduction}
\PARstart{A}{s} one of the key technologies in the fifth generation (5G) wireless communication system, massive multiple-input multiple-output (MIMO) has now motivated a growing research interest \cite{b1}. In massive MIMO systems, hundreds of antenna elements are deployed at the base station (BS). Combined with a pre-coding scheme, such as minimum mean-squared error (MMSE), these antennas provide an effective way to exploit the spatial degrees of freedom, which significantly enhance system performance, e.g., system capacity, energy efficiency, and link robustness \cite{b2}--\cite{b8}.

In massive MIMO systems, the accurate channel state information (CSI) is required by BSs for downlink beamforming user selection \cite{b9}. In the time division duplex (TDD) mode, the CSI of downlink can be estimated by the uplink channel for the reciprocity property \cite{b10}. However, in the frequency division duplex (FDD) mode, the reciprocity-based CSI is not available. Thus, the downlink CSI should be estimated by users and fed back to the BS. This CSI feedback incurs significant overhead in massive MIMO systems due to large number of antennas. Since FDD mode is pervasively deployed for delay sensitive and traffic symmetric applications, it is of great importance to reduce the CSI feedback overhead in FDD mode.

The codebook-based CSI feedback has been widely applied \cite{b11}. In FDD massive MIMO systems, however, the large number of antennas requires correspondingly expanded codebook size to guarantee an acceptable CSI-accuracy \cite{b12}. Subject to the curse of dimensionality, the overhead of the codebook-based feedback becomes substantial for massive MIMO systems \cite{b13}--\cite{b15}. To address the aforementioned problems, the compressive sensing (CS)-based CSI feedback approaches are proposed to reduce the channel dimension by exploiting the sparse structures of CSI \cite{b12},\cite{b14}--\cite{b16} (e.g., CSI's temporal correlation \cite{b12}, CSI's spatial correlation \cite{b14}--\cite{b16}, and the sparsity-enhancing basis for CSI \cite{b14}, etc.). It is well known that, the sparsity of CSI is only approximated for specific models \cite{b3}, \cite{b4}, beyond which, the general assumption of channel sparsity could not be guaranteed. Thus, existing CS-based algorithms may have practical issues in case of model mismatch.

Recently, the deep learning (DL) based physical-layer technique shows its promising prospects in wireless communication system \cite{b3}--\cite{b9}, \cite{b17}--\cite{b21} and the comprehensive overview could be found in \cite{b18}--\cite{b20}. Compared with the CS-based CSI feedback, DL-based methods (e.g., \cite{b3}, and \cite{b4}) outperform many existing CS schemes in feedback reduction. Despite all this, an efficient DL-based CSI feedback to further improve the occupation of the uplink bandwidth resource is still highly desired.

\subsection{RELATED WORKS}

The literature of DL-based CSI feedback for FDD massive MIMO systems mainly concentrates on feedback reduction \cite{b3}--\cite{b6}. In \cite{b3}, a deep neural network (DNN) called CsiNet has been developed for CSI feedback. The CsiNet is based on autoencoder of DNN, where the encoder learns to compress the original channel matrices to some codewords and the decoder learns the inverse transformation from compressed codewords through training data. Compared to the CS-based algorithms, the CsiNet was more effective in reducing the CSI dimensionality. However, the CSI is independently reconstructed in CsiNet and thus it is not suitable for practical application in time-varying channels due to the ignorance of time correlation. To remedy this defect, a CsiNet-long short-term memory (CsiNet-LSTM) has been proposed in \cite{b4} to enhance recovery quality of CSI by learning spatial structures and time correlation of time-varying massive MIMO channels. However, the investigation in \cite{b5} indicated that both \cite{b3} and \cite{b4} (i.e., CsiNet and CsiNet-LSTM) are not sufficient for tracking the temporal correlations due to the employment of linear fully-connected networks (FCNs) for CSI compression. By incorporating a LSTM module and FCN in a neural network (NN) architecture, the recurrent compression and uncompression modules were formed in \cite{b5} to effectively capture the temporal and frequency correlations of wireless channels. Considering feedback error and feedback delay, a deep autoencoder based CSI feedback was proposed in \cite{b6}. Although the DL-based CSI feedback methods in \cite{b3}--\cite{b6} exhibite excellent performance in feedback reduction, the uplink bandwidth resources are still occupied to some extent.

Without any occupation of uplink bandwidth resources, \cite{b7} and \cite{b8} estimated downlink CSI from uplink CSI by using DL approach. In \cite{b7}, the core idea was that since the same propagating environment was shared for both uplink and downlink channels, the environment information could be applied to downlink channel cases after it was extracted from uplink channel response. Similar to \cite{b7}, a NN-based scheme for extrapolating downlink CSI from observed uplink CSI has been proposed in \cite{b8}, where the underlying physical relation between the downlink and uplink frequency bands was exploited to construct the learning architecture. Need to mention that, the methods in \cite{b7} usually needs to retrain the NN when the environment information changes significantly. For example, for a well-trained equipment, its extracted environment information (e.g., the shapes of buildings, streets and mountains, the materials that objects are made up, etc) from one city would no longer be applicable for another. The method in \cite{b8} will encounter poor CSI recovery performance in the environment of wide band interval between downlink and uplink frequency bands.

Besides the DL-based CSI feedback approaches, the superimposed coding (SC), which is similar the non-orthogonal multiple access scheme \cite{b21}, is also proposed for CSI feedback to avoid the occupation of uplink bandwidth resources \cite{b22}. This is accomplished by spreading and superimposing the downlink CSI on the uplink user data sequences (UL-US) to feed back to BS \cite{b22}. But still, this method is challenged by the difficulties of cancelling the interference between CSI and UL-US.

As a whole, the DL-based and SC-based CSI feedback methods still face huge challenge, which can be summarized as follows:
\begin{itemize}
\item Concentrated on feedback reduction, the DL-based CSI feedback methods, e.g., the methods in \cite{b3}--\cite{b6}, inevitably occupy uplink bandwidth resources.
\item Although the occupation of uplink bandwidth resources can be avoided, the methods that estimate downlink CSI from uplink CSI in \cite{b7} and \cite{b8} usually limit the applications in mobile or wide frequency-band interval environment.
\item The SC-based CSI feedback \cite{b22} can also avoid the occupation of uplink bandwidth resources, while facing with huge challenge to cancel the interference between downlink CSI and UL-US due to the lack of good solutions in previous works.
\end{itemize}

Motivated by DL-based CSI feedback methods, we combine DL technique and SC technique for CSI feedback to overcome these challenges mentioned above.

\subsection{CONTRIBUTIONS}
In this paper, we combine DL technique and SC technique for CSI feedback. The main contributions of our work are summarized as follows:
\begin{itemize}
\item The SC-based CSI feedback (e.g., \cite{b22}) is introduced in user equipment. Therefore, the occupation of uplink bandwidth resource is thoroughly avoided, which is different from the DL-based methods in \cite{b3}--\cite{b6}. In particular, the DL-based methods by using uplink CSI to estimate downlink CSI in \cite{b7} and \cite{b8} are not adopted for a wider application in mobile or wide frequency-band interval environment.

\item A multi-layer NN (i.e., a DNN) is constructed at BS by with the unfolding idea from \cite{b23}--\cite{b25}. Compared to the SC-based CSI feedback \cite{b22} with perfectly known noise variance, this multi-layer NN method improves the performance of downlink CSI recovery without obvious change of bit error rate (BER) of UL-US. Note that the iteration algorithm according to minimum mean-squared error (MMSE) criterion in \cite{b22} requires to know the noise variance. Our unfolded iteration can work well without any knowledge of link noise. That is, both the recovery of downlink CSI and the BER of UL-US are actually improved compared to SC-based CSI feedback in \cite{b22} due to the inevitable estimation errors of noise variance.
\item A subnet-by-subnet method, inspired by layer-by-layer training in \cite{b26}, is exploited to train the designed DNN. This method facilitates the parameter tuning and expedites the convergence rate.

\end{itemize}

The remainder of this paper is structured as follows: In Section II, we present the SC-based CSI feedback to formulate a learning problem. The proposed method, i.e., deep learning for CSI feedback is presented in Section III, and the numerical results are given in IV. Finally, Section V concludes our work.

Notations: Boldface letters are used to denote matrices and column vectors;${\left(\cdot \right)^T}$, ${\left(\cdot \right)^H}$, ${\left(\cdot \right)^\dag}$ and $\mathrm {E}\left\{ \cdot \right\}$ denote the transpose, conjugate transpose, matrix pseudo-inverse, and statistical expectation respectively; ${\mathop{\mathrm {Re}}\nolimits} \left(  \cdot  \right)$ and ${\mathop{\mathrm {Im}}\nolimits} \left(  \cdot  \right)$ denote the real and imaginary parts of a complex number, complex vector or complex matrix; ${{\mathbf{I}}_P}$ is the identity matrix of size $P \times P$; ${\mathrm{BN}}\left(  \cdot  \right)$ denotes the operation of batch normalization; ${\left\|  \cdot  \right\|_2}$ is the Euclidean norm; and ${\bm{0}}$ is the matrix or vector with all zero elements.

\section{PROBLEM FORMULATION}

In this section, the SC-based CSI feedback is first elaborated in II-A, and a \textit{SC-baseline} is also formed for ease of comparison and description. Then, in II-B, based on this baseline, we form a multi-task learning for SC-based CSI feedback.

\subsection{SC-BASED CSI FEEDBACK }
In \cite{b22}, the MIMO system consists of a BS with $N$ antennas and $U$ single-antenna users. The transmitting signal $\mathbf{X}_u $ of user-$u$, $u = 1,2, \cdots ,U$, is denoted as
\begin{equation}\label{EQ1}
{{\mathbf{X}}_u} = \sqrt {\frac{{\rho {E_u}}}{N}} {{\mathbf{H}}_u}{\mathbf{P}}_u^T + \sqrt {\left( {1 - \rho } \right){E_u}} {{\mathbf{D}}_u},
\end{equation}
where, $\rho  \in \left[ {0,1} \right]$ stands for the power proportional coefficient (PPC). For each user-$u$, $E_u$ represents the transmitting power; ${{\mathbf{H}}_u}$ is the $1 \times N$ downlink CSI from BS to user-$u$, whose elements are independent and identically distributed (i.i.d) complex Gaussian variable with zero mean and variance ${1 \mathord{\left/
 {\vphantom {1 N}} \right.
 \kern-\nulldelimiterspace} N}$; ${{\mathbf{P}}_u} \in {\mathbb{R}^{M \times N}}$ is a spreading matrix, satisfying ${\mathbf{P}}_u^T{{\mathbf{P}}_u} = M{\kern 1pt} {\kern 1pt} {{\mathbf{I}}_N}$; ${{\mathbf{D}}_u} \in {\mathbb{C}^{1 \times M}}$ denotes UL-US; and $M$ is the frame length (or UL-US length).

The received signal at BS from user-$u$, denoted as $\mathbf {r}_u$, is given by \cite{b22}
\begin{equation}\label{EQ2}
{{\mathbf{r}}_u} = {{\mathbf{G}}_u}{{\mathbf{X}}_u} + {{\mathbf{N}}_u},
\end{equation}
where, $\mathbf {r}_u$ is $N \times M$ signal block captured from $N$ BS antennas; ${{\mathbf{G}}_u} \in {\mathbb{C}^{N \times 1}}$ is uplink channel vector, i.e., uplink CSI; the feedback link noise is denoted by ${{\mathbf{N}}_u}$, which is a   $N \times M$ complex matrix. Each element of ${{\mathbf{N}}_u}$ is modeled as i.i.d complex additive white Gaussian noise (AWGN) with zero mean and variance $\sigma _u^2$.

Assuming perfect synchronization, perfect uplink channel estimation (i.e., ${{\mathbf{G}}_u}$ can be known), and perfect noise variance estimation (i.e., $\sigma _u^2$ is known) to be available at the BS, we form a ``\textit{SC-baseline}'' for DL-based CSI feedback. Referring to \cite{b22}, the iteration procedure of ``\textit{SC-baseline}'', which is utilized to recover downlink CSI and UL-US on the basis of MMSE criterion, is given as follows:
\begin{enumerate}
\item Initialization: $k=0$, ${\mathbf{r}}_u^{\left( 0 \right)} \leftarrow {{\mathbf{r}}_u}$.

\item MMSE estimation of downlink CSI (i.e., ${\mathbf{\mathord{\buildrel{\lower3pt\hbox{$\scriptscriptstyle\smile$}}
\over H} }}_u^{\left( k \right)}$): Compute ${\mathbf{Z}}_u^{\left( k \right)} = {{{\mathbf{r}}_u^{\left( k \right)}{{\mathbf{P}}_u}} \mathord{\left/
 {\vphantom {{{\mathbf{r}}_u^{\left( k \right)}{{\mathbf{P}}_u}} M}} \right.
 \kern-\nulldelimiterspace} M}$ to despread the updated signal ${\mathbf{r}}_u^{\left( k \right)}$, and then estimate the downlink CSI according to MMSE criterion, i.e.,
 \begin{equation}\label{EQ3}
\begin{array}{c}
{\mathbf{\mathord{\buildrel{\lower3pt\hbox{$\scriptscriptstyle\smile$}}
\over H} }}_u^{\left( k \right)} = M\sqrt {\rho {E_u}N} \left\{ {\left[ {N + \left( {M - N} \right)\rho } \right]} \right. \times \\
{\left. {{\kern 1pt} {\kern 1pt} {\kern 1pt} {\kern 1pt} {\kern 1pt} {\kern 1pt} {E_u}{\mathbf{G}}_u^H{{\mathbf{G}}_u} + N\sigma _u^2} \right\}^{ - 1}}{\mathbf{G}}_u^H{\mathbf{Z}}_u^{\left( k \right)}.
\end{array}
\end{equation}

\item Eliminate the interference of downlink CSI:
\begin{equation}\label{EQ4}
 {\mathbf{r}}_u^{\left( k \right)} \leftarrow {\mathbf{r}}_u^{\left( k \right)} - \sqrt {\frac{{\rho {E_u}}}{N}} {{\mathbf{G}}_u}{\mathbf{\mathord{\buildrel{\lower3pt\hbox{$\scriptscriptstyle\smile$}}
\over H} }}_u^{\left( k \right)}{\mathbf{P}}_u^T.
\end{equation}

\item MMSE detection of UL-US (i.e., ${\mathbf{\mathord{\buildrel{\lower3pt\hbox{$\scriptscriptstyle\smile$}}
\over D} }}_u^{\left( k \right)}$)
\begin{equation}\label{EQ5}
\begin{array}{l}
{\mathbf{\mathord{\buildrel{\lower3pt\hbox{$\scriptscriptstyle\smile$}}
\over D} }}_u^{\left( k \right)} = \sqrt {(1 - \rho ){E_u}} \left\{ {(1 - \rho ){E_u}{\mathbf{G}}_u^H{{\mathbf{G}}_u} + } \right.\\
{~~~~~~~~~~~~~\left. {\sigma _u^2} \right\}^{ - 1}} \times {\mathbf{G}}_u^H{\mathbf{r}}_u^{\left( k \right)}
\end{array}
\end{equation}

\item Cancellation of UL-US's interference:
 \begin{equation}\label{EQ6}
{\mathbf{r}}_u^{\left( k \right)} \leftarrow {\mathbf{r}}_u^{\left( k \right)} - \sqrt {\left( {1 - \rho } \right){E_u}} {{\mathbf{G}}_u}{\mathbf{\mathord{\buildrel{\lower3pt\hbox{$\scriptscriptstyle\smile$}}
\over D} }}_u^{\left( k \right)}.
\end{equation}

\item $k = k + 1$ and return to step 2) if $k$ is within iteration limit.

 \end{enumerate}
It should be noted that, to form a comparison baseline, the maximum likelihood detection of UL-US and maximum likelihood estimation of downlink CSI, is impractical due to the extremely high computational complexity in a massive MIMO system. Therefore, the MMSE criterion is considered here for \textit{SC-baseline}. After several iterations, the MMSE estimation of downlink CSI and the MMSE detection of UL-US could be converged.

\Figure[t!](topskip=0pt, botskip=0pt, midskip=0pt){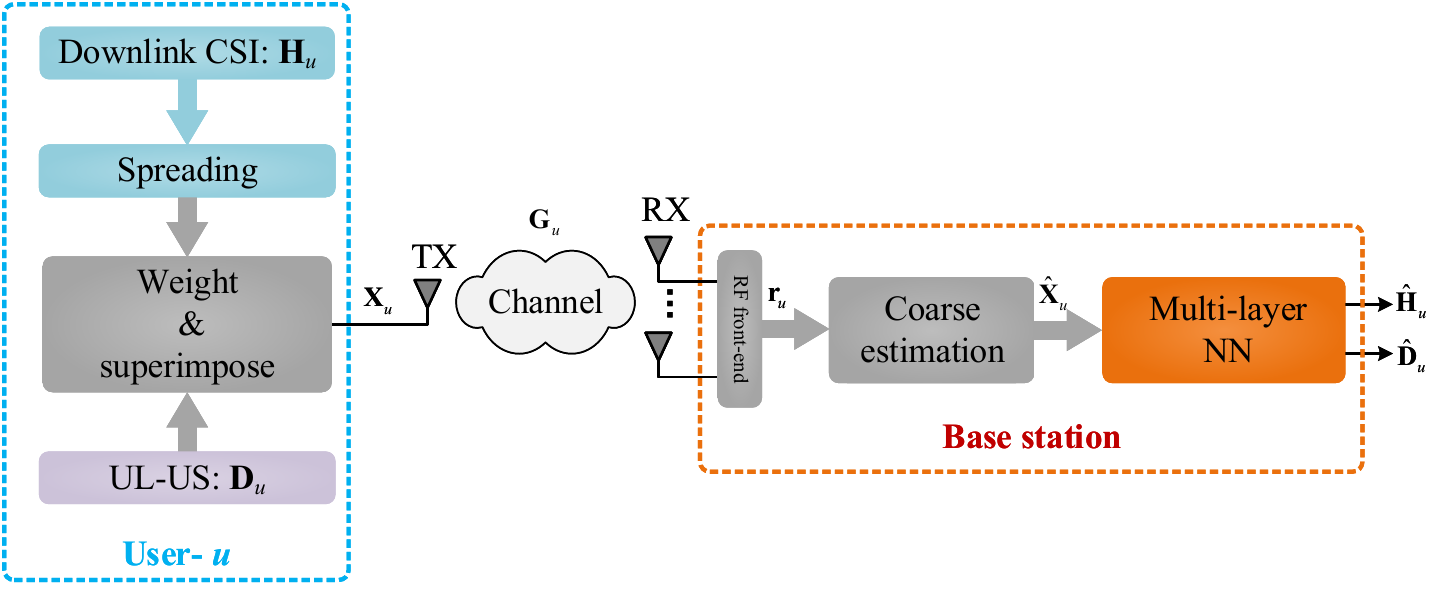}
{System model of the multi-task learning for SC-based CSI feedback.\label{fig1}}

\subsection{ LEARNING TASK FOR SC-BASED CSI FEEDBACK}
To further improve the SC-based CSI feedback, we combine the DL and SC for CSI feedback by exploiting the advantages of SC and DL techniques.
The whole system model is given in Fig.~\ref{fig1}. For user-$u$, the downlink CSI (i.e., ${{\mathbf{H}}_u}$) is spread firstly. Then the weighted downlink CSI and UL-US are superimposed together to form signal ${{\mathbf{X}}_u}$, as given in (\ref{EQ1}). Over the attenuation of the uplink channel ${{\mathbf{G}}_u}$ and link noise ${{\mathbf{N}}_u}$, the transmitted ${{\mathbf{X}}_u}$ from user-$u$ is received at BS. Experiencing the operation of radio frequency (RF) front-end, the received signal ${{\mathbf{r}}_u}$ is expressed in (\ref{EQ2}). With the received signal ${{\mathbf{r}}_u}$, the main task of BS is to recover downlink CSI and detect UL-US by using DL technique.

Similar to the assumption of \cite{b22} and \cite{b24}, the uplink channel ${{\mathbf{G}}_u}$ (i.e., the uplink CSI) is known to the BS in advance. In \cite{b24}, the knowledge of CSI is used to form maximum likelihood optimization for DL-based MIMO detection problem. However, the complicated NN architecture (e.g., 30 layers in \cite{b24}), long training time (e.g., 3 days in \cite{b24}), and difficult parameter tuning, etc., cause its application difficulties in different scenarios. Besides the detection of UL-US (i.e., ${{\mathbf{D}}_u}$), the estimation of downlink CSI (i.e., ${{\mathbf{H}}_u}$) is also needed at the BS. This is a typical multi-task problem in NN \cite{b27}, which encounters more difficulties than the single-task detection (e.g., \cite{b24}). Therefore, to simplify implementation complexity, a multi-task NN architecture is structured by unfolding the iterations of SC-baseline under MMSE criterion. Naturally, other baselines and corresponding NN architectures formed according to the same approach can also be considered, which will not affect the fairness of the comparison.

Although the known uplink CSI ${{\mathbf{G}}_u}$ is exploited in SC-baseline under MMSE criterion, we are still trying to develop a multi-task NN that has no uplink CSI as input but outperforms SC-baseline. Thus, a coarse estimation of ${{\mathbf{X}}_u}$ is employed to circumvent the explicit uplink CSI  ${{\mathbf{G}}_u}$. To do this, the NN architecture can be simplified and thus accelerates network convergence. Then, the estimated ${{\mathbf{\hat X}}_u}$ passes through a multi-layer NN (i.e., a DNN) to solve the multi-task problem, i.e., to recover downlink CSI (denoted as ${{\mathbf{\hat H}}_u}$) and to detect UL-US (denoted as ${{\mathbf{\hat D}}_u}$). This will be elaborated in the next section.

\section{DEEP LEARNING FOR CSI FEEDBACK}
In traditional SC-based CSI feedback \cite{b22}, the main task of BS is to recover downlink CSI and detect UL-US. In our proposed DL-based CSI solution, this is also the main task at BS. From II-B, a coarse estimation is employed for simplification and convergence acceleration of designed DNN. In this section, the coarse estimation is first described and then followed by our multi-layer NN design, in which the downlink CSI recovery and UL-US detection is addressed by solving a multi-task problem.

\subsection{COARSE ESTIMATION}
The benefit of a coarse estimation is to eliminate the interference of uplink channel. When the uplink CSI is not used as network input, the NN architecture can be simplified, and thus improves the convergence rate of offline training. According to the received signal ${{\mathbf{r}}_u}$ at BS, the coarse estimation can be given by
 \begin{equation}\label{EQ7}
{{\mathbf{\hat X}}_u} = {\mathbf{G}}_u^\dag {{\mathbf{r}}_u} = {{\mathbf{X}}_u} + {\mathbf{G}}_u^\dag {{\mathbf{N}}_u}.
\end{equation}
Then, the estimated ${{\mathbf{\hat X}}_u}$ is delivered to a multi-layer NN, and a multi-task problem is solved in the next subsection.

\Figure[t!](topskip=0pt, botskip=0pt, midskip=0pt){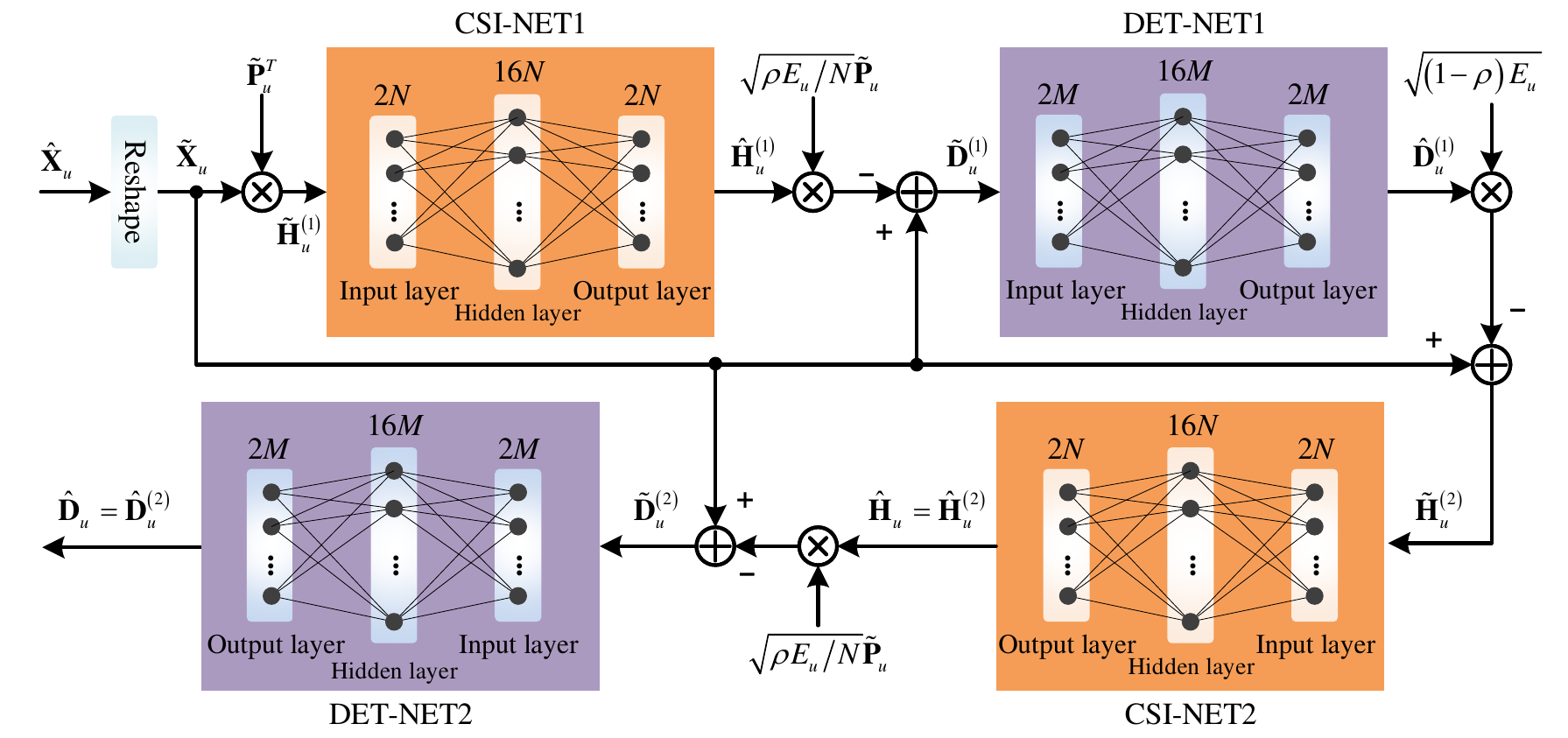}
{Multi-layer NN architecture.\label{fig2}}

\subsection{MULTI-TASK DL NETWORK}
To solve our multi-task problem (i.e., to recover downlink CSI ${{\mathbf{H}}_u}$ and to detect UL-US ${{\mathbf{D}}_u}$), a multi-layer NN is constructed by unfolding the iteration of SC-baseline in II-A. In \cite{b22}, simulations show that with three iterations, the SC-based feedback algorithm nearly converges. According to our design and experiment, we observed that unfolding two iterations is enough. Unfolding with more iterations could not obtain significant improvement to recover downlink CSI and UL-US but merely increase the complexity of NN. Thus, without special explanation, the unfolding operation in the rest of this paper is applied on a two iterations' SC-baseline, and this forms a four subnets' NN. Need to mention that, this subnet structure is flexible for unfolding three or more iterations. The designed multi-layer NN is illuminated in Fig.~\ref{fig2}.

\subsubsection{NETWORK FUNCTION SUMMARY}
For ease of description, we denote four subnets as CSI-NET1, DET-NET1, CSI-NET2, and DET-NET2, respectively. The functionality of the network components is summarized as follows:
\begin{itemize}
\item CSI-NET$i$ corresponds to the MMSE estimation of downlink CSI (i.e., (\ref{EQ3}) in SC-baseline), while $i=1,2$ represents the first and second iteration, respectively.
\item DET-NET1 and DET-NET2 respectively detect UL-US (i.e., (\ref{EQ5}) in SC-baseline) in the first and second iteration.
\item Some known parameters and iteration procedure, corresponding to (\ref{EQ4}) and (\ref{EQ6}) in SC-baseline, are exploited as expert knowledge to implement interference reduction. In addition, this expert knowledge is also utilized to improve network performance, e.g., the convergence acceleration \cite{b28}.
\end{itemize}

\subsubsection{NETWORK ARCHITECTURE}
In Fig.~\ref{fig2}, each of the four subnets consists of an input layer, a hidden layer, and an output layer with a fully connected (FC) mode. These subnets look straightforward, but they are very conducive to parameter tuning in III-C. The architecture is given as follows:
\begin{itemize}
\item CSI-NET1, DET-NET1, CSI-NET2, and DET-NET2 are successively cascaded to form a multi-task network. In addition, some expert knowledge is inserted between two cascaded subnets to implement interference reduction.
\item For CSI-NET1 or CSI-NET2 (DET-NET1 or DET-NET2), the neuron numbers of input layer, hidden layer, and output layer are $2N$ ($2M$), $16N$ ($16M$), and $2N$ ($2M$), respectively.
\item For each subnet, the batch normalization (BN), which is used to accelerate convergence and prevent overfitting \cite{b29}, is employed to normalize input layer and hidden layer. To do so, the inputs of these layers will have zero mean and unit variance.
\item For each subnet, the hidden layer adopts activation function ``swish'', defined as ${\mathrm{swish}}\left( x \right){{ = x} \mathord{\left/
 {\vphantom {{ = x} {\left( {1 + {e^{ - x}}} \right)}}} \right.
 \kern-\nulldelimiterspace} {\left( {1 + {e^{ - x}}} \right)}}$, for a usual good performance \cite{b30} \cite{b31}. Linear activation is employed for other layers which are not listed here.
\item The outputs of CSI-NET2 and DET-NET2 are the estimated downlink CSI ${{\mathbf{\hat H}}_u}$ and detected UL-US ${{\mathbf{\hat D}}_u}$, respectively.

\end{itemize}

\subsubsection{NETWORK PROCESSING}

\begin{itemize}
  \item \textit{Data Preprocessing}
\end{itemize}

In the common framework of machine learning, the data set has to be real value. However, signals in wireless systems are complex valued. Thus, to make the NN architecture in Fig.~\ref{fig2} works, the data preprocessing is first given. The complex vectors of downlink CSI ${{\mathbf{H}}_u} \in {\mathbb{C}^{1 \times N}}$, UL-US ${{\mathbf{D}}_u} \in {\mathbb{C}^{1 \times M}}$ and estimated ${{\mathbf{\hat X}}_u} \in {\mathbb{C}^{1 \times M}}$ (see the coarse estimation in III-A) are reshaped as real valued vectors ${{\mathbf{\tilde H}}_u} \in {\mathbb{R}^{2N \times 1}}$, ${{\mathbf{\tilde D}}_u} \in {\mathbb{R}^{2M \times 1}}$ and ${{\mathbf{\tilde X}}_u} \in {\mathbb{R}^{2M \times 1}}$, respectively, i.e.,
 \begin{equation}\label{EQ8}
{{\mathbf{\tilde H}}_u} = {\left[ {{\mathop{\mathrm {Re}}\nolimits} \left( {{{\mathbf{H}}_u}} \right),{\kern 1pt} {\kern 1pt} {\kern 1pt} {\kern 1pt} {\kern 1pt} {\mathop{\mathrm {Im}}\nolimits} \left( {{{\mathbf{H}}_u}} \right)} \right]^T},
\end{equation}
 \begin{equation}\label{EQ9}
{{\mathbf{\tilde D}}_u} = {\left[ {{\mathop{\mathrm {Re}}\nolimits} \left( {{{\mathbf{D}}_u}} \right),{\kern 1pt} {\kern 1pt} {\kern 1pt} {\kern 1pt} {\kern 1pt} {\mathop{\mathrm {Im}}\nolimits} \left( {{{\mathbf{D}}_u}} \right)} \right]^T},
\end{equation}
 \begin{equation}\label{EQ10}
{{\mathbf{\tilde X}}_u}{\mathrm{ = }}{\left[ {{\mathop{\mathrm {Re}}\nolimits} \left( {{{{\mathbf{\hat X}}}_u}} \right),{\kern 1pt} {\kern 1pt} {\kern 1pt} {\kern 1pt} {\kern 1pt} {\mathop{\mathrm {Im}}\nolimits} \left( {{{{\mathbf{\hat X}}}_u}} \right)} \right]^T}.
\end{equation}
To match real valued vectors operation, we also transform the spreading matrix ${{\mathbf{P}}_u} \in {\mathbb{R}^{M \times N}}$ as
 \begin{equation}\label{EQ11}
{{\mathbf{\tilde P}}_u} = \left[ {\begin{array}{*{20}{c}}
{{{\mathbf{P}}_u}}&{\mathbf{0}}\\
{\mathbf{0}}&{{{\mathbf{P}}_u}}
\end{array}} \right].
\end{equation}
Then, the reshaped real valued vector ${{\mathbf{\tilde X}}_u}$ is used as the input of the process in TABLE~\ref{table_I}.

\begin{table}[!ht]
\renewcommand\arraystretch{1.2}
\caption{PROCESSING PROCEDURE}
\label{table_I}
\begin{tabular}{l}
\hline
\hline
\\
\textbf{Input:} ${{\mathbf{\tilde X}}_u} \in {\mathbb{R}^{2M \times 1}}$ \\\\

\hline \\
\kern 9pt (0-1): Despread: ${\mathbf{\tilde H}}_u^{\left( 1 \right)} = {\mathbf{\tilde P}}_u^T{{\mathbf{\tilde X}}_u}.$\\

\kern 9pt (1-1): Use CSI-NET1 to estimate downlink CSI, then we gain ${\mathbf{\hat H}}_u^{\left( 1 \right)}$.\\

\kern 9pt (1-2): Reduce downlink CSI interference with the expert \\
   \kern 30pt knowledge to obtain ${\mathbf{\tilde D}}_u^{\left( 1 \right)}$.\\

\kern 9pt (1-3): Detect UL-US by using DET-NET1 to acquire $
{\mathbf{\hat D}}_u^{\left( 1 \right)}$.\\

\kern 9pt (1-4): Diminish UL-US interference by using expert knowledge \\
    \kern 30pt  to capture ${\mathbf{\tilde H}}_u^{\left( 2 \right)}.$\\

\kern 9pt (2-1): Employ CSI-NET2 to estimate downlink CSI and acquire ${\mathbf{\hat H}}_u^{\left( 2 \right)}$.\\

\kern 9pt (2-2): Decrease downlink CSI interference by using the expert\\
   \kern 30ptknowledge to obtain ${\mathbf{\tilde D}}_u^{\left( 2 \right)}$.\\

\kern 9pt (2-3): Detect UL-US by using DET-NET2 to achieve ${\mathbf{\hat D}}_u^{\left( 2 \right)}$.\\\\
\hline \\

\textbf{Output:} ${{\mathbf{\hat H}}_u} = {\mathbf{\hat H}}_u^{\left( 2 \right)}$ and ${{\mathbf{\hat D}}_u} = {\mathbf{\hat D}}_u^{\left( 2 \right)}$.\\
 \\
 \hline
 \hline
\end{tabular}
\end{table}

\begin{itemize}
  \item \textit{Processing Procedure}
\end{itemize}

  The procedure of proposed NN is given in TABLE \ref{table_I}, and some steps are explained as follows. For the sake of convenience, we use ${{\mathbf{W}}_{X1}}$ (${{\mathbf{b}}_{X1}}$) to denote the weight matrices (bias vectors) for hidden layer, while  and ${{\mathbf{W}}_{X2}}$ (${{\mathbf{b}}_{X2}}$) for output layer, respectively. Where $X = Ci$ or $Di$ represent the CSI-NET$i$ and DET-NET$i$, $i=1,2$, respectively.

\textbf{Despreading}: With the mapped real valued vector  ${{\mathbf{\tilde X}}_u}$, a despreading (see (0-1) in TABLE \ref{table_I}) is employed to reduce UL-US interference. The corresponding despreading at BS can be expressed as
\begin{equation}\label{EQ12}
{\mathbf{\tilde H}}_u^{\left( 1 \right)} = {\mathbf{\tilde P}}_u^T{{\mathbf{\tilde X}}_u},
\end{equation}
where ${\mathbf{\tilde P}}_u^T$ is obtained by transforming ${\mathbf{P}}_u$ according to (\ref{EQ11}). The despreading is used to reduce UL-US interference, which is corresponded to the despreading in (\ref{EQ3}).

\textbf{Estimation of downlink CSI}: The step (1-1) and (2-1) in TABLE~\ref{table_I} are used to estimate downlink CSI according to CSI-NET1 and CSI-NET2, respectively. These estimations can be given by
\begin{equation}\label{EQ13}
\begin{array}{l}
{\mathbf{\hat H}}_u^{\left( i \right)} = {{\mathbf{W}}_{Ci2}}{\mathrm{BN}}\left( {{\mathrm{swish}}\left( {{{\mathbf{W}}_{Ci1}}{\mathrm{BN}}\left( {{\mathbf{\tilde H}}_u^{\left( i \right)}} \right) + {{\mathbf{b}}_{Ci1}}} \right)} \right)\\
 ~~~~~~~~~~~~+ {{\mathbf{b}}_{Ci2}}.
\end{array}
\end{equation}
Where ${{\mathbf{W}}_{Ci1}} \in {\mathbb{R}^{16N \times 2N}}$, ${{\mathbf{W}}_{Ci2}} \in {\mathbb{R}^{2N \times 16N}}$, ${{\mathbf{b}}_{Ci1}} \in {\mathbb{R}^{16N \times 1}}$ and ${{\mathbf{b}}_{Ci2}} \in {\mathbb{R}^{2N \times 1}}$. The operations in (\ref{EQ13}) correspond to the MMSE estimation of downlink CSI of the $i$th iteration in (\ref{EQ3}).

\textbf{Reduction of downlink CSI interference}: We use the step (1-2) and (2-2) in TABLE \ref{table_I} to reduce the downlink CSI interference. According to ${\mathbf{\hat H}}_u^{\left( i \right)}$, ${{\mathbf{\tilde X}}_u}$, and the expert knowledge, the interference reduction can be given by
\begin{equation}\label{EQ14}
{\mathbf{\tilde D}}_u^{\left( i \right)} = {{\mathbf{\tilde X}}_u} - \sqrt {\frac{{\rho {E_u}}}{N}} {{\mathbf{\tilde P}}_u}{\mathbf{\hat H}}_u^{\left( i \right)},{\kern 1pt} {\kern 1pt} {\kern 1pt} i = 1,2.
\end{equation}
where the known ${{\mathbf{\tilde P}}_u}$, $E_u$, $\rho$, $N$ and the structure of interference reduction are viewed as \textit{expert knowledge}. These interference reductions are related to the $i$th iteration in (\ref{EQ4}).

\textbf{Detection of UL-US}: The UL-US detections are given in step (1-3) and (2-3) based on DET-NET1 and DET-NET2, respectively. The detection can be expressed as
\begin{equation}\label{EQ15}
\begin{array}{l}
{\mathbf{\hat D}}_u^{\left( i \right)} = {{\mathbf{W}}_{Di2}}{\mathrm{BN}}\left( {{\mathrm{swish}}\left( {{{\mathbf{W}}_{Di1}}{\mathrm{BN}}\left( {{\mathbf{\tilde D}}_u^{\left( i \right)}} \right) + {{\mathbf{b}}_{Di1}}} \right)} \right)\\
 ~~~~~~~~~~~~+ {{\mathbf{b}}_{Di2}}.
\end{array}
\end{equation}
where ${{\mathbf{W}}_{Di1}} \in {\mathbb{R}^{16M \times 2M}}$, ${{\mathbf{W}}_{Di2}} \in {\mathbb{R}^{2M \times 16M}}$, ${{\mathbf{b}}_{Di1}} \in {\mathbb{R}^{16M \times 1}}$ and ${{\mathbf{b}}_{Di2}} \in {\mathbb{R}^{2M \times 1}}$. In (\ref{EQ15}), the detection is related to the MMSE detection of UL-US of  $i$th iteration in (\ref{EQ5}).

\textbf{UL-US interference reduction}: In TABLE~\ref{table_I}, the step (1-4) is used to reduce the UL-US interference, which can be given by
\begin{equation}\label{EQ16}
{\mathbf{\tilde H}}_u^{\left( 2 \right)} = {{\mathbf{\tilde X}}_u} - \sqrt {\left( {1 - \rho } \right){E_u}} {\mathbf{\hat D}}_u^{\left( 1 \right)},
\end{equation}
where $E_u$, $\rho$, and the structure of interference reduction are known as expert knowledge. This step is corresponded to the interference reduction in (\ref{EQ6}).

By the end of our multi-task network, ${{\mathbf{\hat H}}_u} = {\mathbf{\hat H}}_u^{\left( 2 \right)}$ and ${{\mathbf{\hat D}}_u} = {\mathbf{\hat D}}_u^{\left( 2 \right)}$, or say the outputs of CSI-NET2 and DET-NET2, are the ultimate outputs of downlink CSI estimation and UL-US detection, respectively.

\subsection{MODEL TRAINING SPECIFICATION}
Training a multi-task deep network is usually challenged by vanishing gradient, initialization sensitivity, activation saturation, and model over-fitting \cite{b24}, \cite{b32}, \cite{b33}, \cite{b34}, etc. To overcome these challenges, the common method is to solve an optimization problem by using the gradients of each task to update the shared parameters \cite{b33}. However, the task imbalances impede proper training \cite{b34}, and result in enormous difficulties for parameter tuning.

\subsubsection{SUBNET-BY-SUBNET TRAINING}
To address the challenge of paramter tuning, we come up with a subnet-by-subnet training pattern inspired by the layer-by-layer training in \cite{b26}. Specifically, CSI-NET1 is first trained independently until it converges. Then the weight matrices and bias vectors of CSI-NET1 are fixed and applied to train the next subnet in sequence, i.e., DET-NET1, CSI-NET2 and DET-NET2. The detailed training procedure is given in TABLE~\ref{table_II}.

\begin{table}[!ht]
\renewcommand\arraystretch{1.2}
\caption{SUBNET-BY-SUBNET TRAINING}
\label{table_II}
\setlength{\tabcolsep}{3pt}
\begin{tabular}{l}
\hline
\hline
\\
1. Train CSI-NET1 to obtain the weight matrices (${{\mathbf{W}}_{C11}}$ and ${{\mathbf{W}}_{C12}}$) \\
 \kern 8pt and bias vectors (${{\mathbf{b}}_{C11}}$ and ${{\mathbf{b}}_{C12}}$).\\
2. Maintaining $\left\{ {{{\mathbf{W}}_{C11}},{{\mathbf{W}}_{C12}},{{\mathbf{b}}_{C11}},{{\mathbf{b}}_{C12}}} \right\}$ unchanged, train \\
\kern 8pt DET-NET1, and obtain the weight matrices (${{\mathbf{W}}_{D11}}$ and ${{\mathbf{W}}_{D12}}$) \\ \kern 8pt and bias vectors (${{\mathbf{b}}_{D11}}$ and ${{\mathbf{b}}_{D12}}$).\\

3. Keeping $\left\{ {{{\mathbf{W}}_{C11}},{{\mathbf{W}}_{C12}},{{\mathbf{b}}_{C11}},{{\mathbf{b}}_{C12}},{{\mathbf{W}}_{D11}},{{\mathbf{W}}_{D12}},{{\mathbf{b}}_{D11}},} \right.$\\
\kern 8pt $\left. {{{\mathbf{b}}_{D12}}} \right\}$ unchanged, we train CSI-NET2 to acquire the weight \\
\kern 8pt matrices (${{\mathbf{W}}_{C21}}$ and ${{\mathbf{W}}_{C22}}$) and bias vectors (${{\mathbf{b}}_{C21}}$ and ${{\mathbf{b}}_{C22}}$).\\

4. Retaining $\left\{ {{{\mathbf{W}}_{C11}},{{\mathbf{W}}_{C12}},{{\mathbf{b}}_{C11}},{{\mathbf{b}}_{C12}},{{\mathbf{W}}_{D11}},{{\mathbf{W}}_{D12}},{{\mathbf{b}}_{D11}},} \right.$ \\
\kern 8pt ${{\mathbf{b}}_{D12}},{{\mathbf{W}}_{C21}},{{\mathbf{W}}_{C22}},\left. {{{\mathbf{b}}_{C21}},{{\mathbf{b}}_{C22}}} \right\}$ unchanged, train \\
\kern 8pt DET-NET2 to achieve the weight matrices (${{\mathbf{W}}_{D21}}$ and ${{\mathbf{W}}_{D22}}$)\\
\kern 8pt and bias vectors (${{\mathbf{b}}_{D21}}$ and ${{\mathbf{b}}_{D22}}$).\\
5. Save $\left\{ {{{\mathbf{W}}_{Ci1}},{{\mathbf{W}}_{Ci2}},{{\mathbf{b}}_{Ci1}},{{\mathbf{b}}_{Ci2}},{{\mathbf{W}}_{Di1}},{{\mathbf{W}}_{Di2}},{{\mathbf{b}}_{Di1}},{{\mathbf{b}}_{Di2}}} \right\}_{i = 1}^2$ \\
\kern 8pt for testing.\\ \\

 \hline
 \hline
\end{tabular}
\end{table}

In the following paragraphs, we first give loss functions involved in training. Then, the initialization of weight matrices and bias vectors are presented. Finally, we explain how to prepare training data.

\subsubsection{LOSS FUNCTIONS}
To train each subnet, the criterion of minimizing the mean squared error (MSE) is used. The loss function for CSI-NET$i$ is expressed as
\begin{equation}\label{EQ17}
Los{s_{{\textrm{CSI-NET}}i}} = \frac{1}{{{T_{1,i}}}}\sum\limits_{t = 1}^{{T_{1,i}}} {\left\| {{{{\mathbf{\tilde H}}}_u} - {\mathbf{\hat H}}_u^{\left( i \right)}} \right\|_2^2} ,{\kern 1pt} {\kern 1pt} {\kern 1pt} {\kern 1pt} {\kern 1pt} {\kern 1pt} i = 1,2,
\end{equation}
where $T_{1,i}$ is the total number of samples in training set of CSI-NET$i$ training, ${{\mathbf{\tilde H}}_u}$ is the real representation of complex vector ${{\mathbf{ H}}_u}$ (see (\ref{EQ11})). Similarly, the loss function for DET-NET$i$ can be given by
\begin{equation}\label{EQ18}
Los{s_{{\textrm{DET-NET}}i}} = \frac{1}{{{T_{2,i}}}}\sum\limits_{j = 1}^{{T_{2,i}}} {\left\| {{{{\mathbf{\tilde D}}}_u} - {\mathbf{\hat D}}_u^{\left( i \right)}} \right\|_2^2} ,{\kern 1pt} {\kern 1pt} {\kern 1pt} {\kern 1pt} {\kern 1pt} {\kern 1pt} i = 1,2,
\end{equation}
where $T_{2,i}$ is the total number of samples in the training set of DET-NET$i$ training.

\subsubsection{WEIGHT AND BIAS INITIALIZATION}
Appropriate initialization can effectively avoid gradient exploding or vanishing problem \cite{b35}. Thus, the initialization of weight matrices and bias vectors should be carefully considered. In this paper, we initialize weight matrices on the basis of the method in \cite{b35}.

For the training of CSI-NET$i$ ($i=1,2$), elements of ${{\mathbf{W}}_{Ci1}}$ and ${{\mathbf{W}}_{Ci2}}$, are initialized as the i.i.d. Gaussian distribution with 0 mean and variance $1/(8N)$ and $1/N$, respectively. Similarly, for the training of DET-NET$i$, elements of ${{\mathbf{W}}_{Di1}}$ and ${{\mathbf{W}}_{Di2}}$ are initialized as the i.i.d. Gaussian distribution with 0 mean and variance $1/(8M)$ and $1/M$, respectively.
Elements of all bias vectors (i.e., ${{\mathbf{b}}_{Ci1}}$, ${{\mathbf{b}}_{Ci2}}$, ${{\mathbf{b}}_{Di1}}$, and ${{\mathbf{b}}_{Di2}}$) are initialized as zeros.

\subsubsection{DATA PREPARATION FOR TRAINING}
The training set is acquired by a simulation approach, in which significant amount of data samples are generated to train a DNN. Specially, these data samples are generated as follows.

${{\mathbf{P}}_u}$ consists of $N$ Walsh codes of length $M$, satisfying ${\mathbf{P}}_u^T{{\mathbf{P}}_u} = M{\kern 1pt} {\kern 1pt} {{\mathbf{I}}_N}$; and ${{\mathbf{\tilde P}}_u}$ is obtained from ${{\mathbf{P}}_u}$ according to (\ref{EQ11}). ${{\mathbf{H}}_u}$ and ${{\mathbf{G}}_u}$ are randomly generated on the basis of the distribution $\mathcal{CN}\left( {0,\left( {{1 \mathord{\left/
 {\vphantom {1 N}} \right.
 \kern-\nulldelimiterspace} N}} \right){{\mathbf{I}}_N}} \right)$. Then complex valued ${{\mathbf{H}}_u}$ is converted to a real valued ${{\mathbf{\tilde H}}_u}$ by using (\ref{EQ8}). The uplink and downlink channels (i.e., ${{\mathbf{H}}_u}$ and ${{\mathbf{G}}_u}$) are assumed to be stable during one frame, but varying from one to another \cite{b36} \cite{b37}. Elements of link noise ${{\mathbf{N}}_u}$ follow the distribution of $ \mathcal{CN}\left( {0,\sigma _u^2} \right)$. $\left\{ {{{\mathbf{D}}_u}} \right\}$ is created by quadrature-phase-shift-keying (QPSK) symbol set generated by modulating a Bernoulli sequence $\left\{ {{s_j}} \right\}$, and then are mapped to $\left\{ {{{{\mathbf{\tilde D}}}_u}} \right\}$ according to (\ref{EQ9}). By using $\left\{ {{{\mathbf{H}}_u}} \right\}$, $\left\{ {{{\mathbf{D}}_u}} \right\}$, $\left\{ {{{\mathbf{G}}_u}} \right\}$ and $\left\{ {{{\mathbf{N}}_u}} \right\}$, we derive training data sets $\left\{ {{{{\mathbf{\tilde X}}}_u}} \right\}$ according to (\ref{EQ1}), (\ref{EQ2}), (\ref{EQ7}) and (\ref{EQ10}). The training labels of estimating $\left\{ {{{\mathbf{H}}_u}} \right\}$ in CSI-NET1 and CSI-NET2 are set as $\left\{ {{{{\mathbf{\tilde H}}}_u}} \right\}$. To detect $\left\{ {{{\mathbf{D}}_u}} \right\}$, the labels used for training DET-NET1 and DET-NET2 are set as $\left\{ {{{{\mathbf{\tilde D}}}_u}} \right\}$.
 \

\section{SIMULATION RESULTS}
In this section, the performance comparison is made between the proposed DL-based scheme and SC-baseline~\cite{b22} (presented in II-A) under different conditions. Some definitions involved in simulations are first given as following. The signal-to-noise ratio (SNR) in decibel (dB) of the received signal from user-$u$ at BS is defined as
\begin{equation}\label{EQ19}
SNR = 10\log_{10}\left( {\frac{{{E_u}}}{{\sigma _u^2}}} \right).
\end{equation}
Normalized MSE (NMSE) is used to evaluate the recovery of downlink CSI, which is defined as
\begin{equation}\label{EQ20}
NMSE = {\mathrm{E}}\left\{ {\frac{{\left\| {{{{\mathbf{\tilde H}}}_u} - {{{\mathbf{\hat H}}}_u}} \right\|_2^2}}{{\left\| {{{{\mathbf{\tilde H}}}_u}} \right\|_2^2}}} \right\}.
\end{equation}

\begin{figure}
\centering
\includegraphics[scale=0.75]{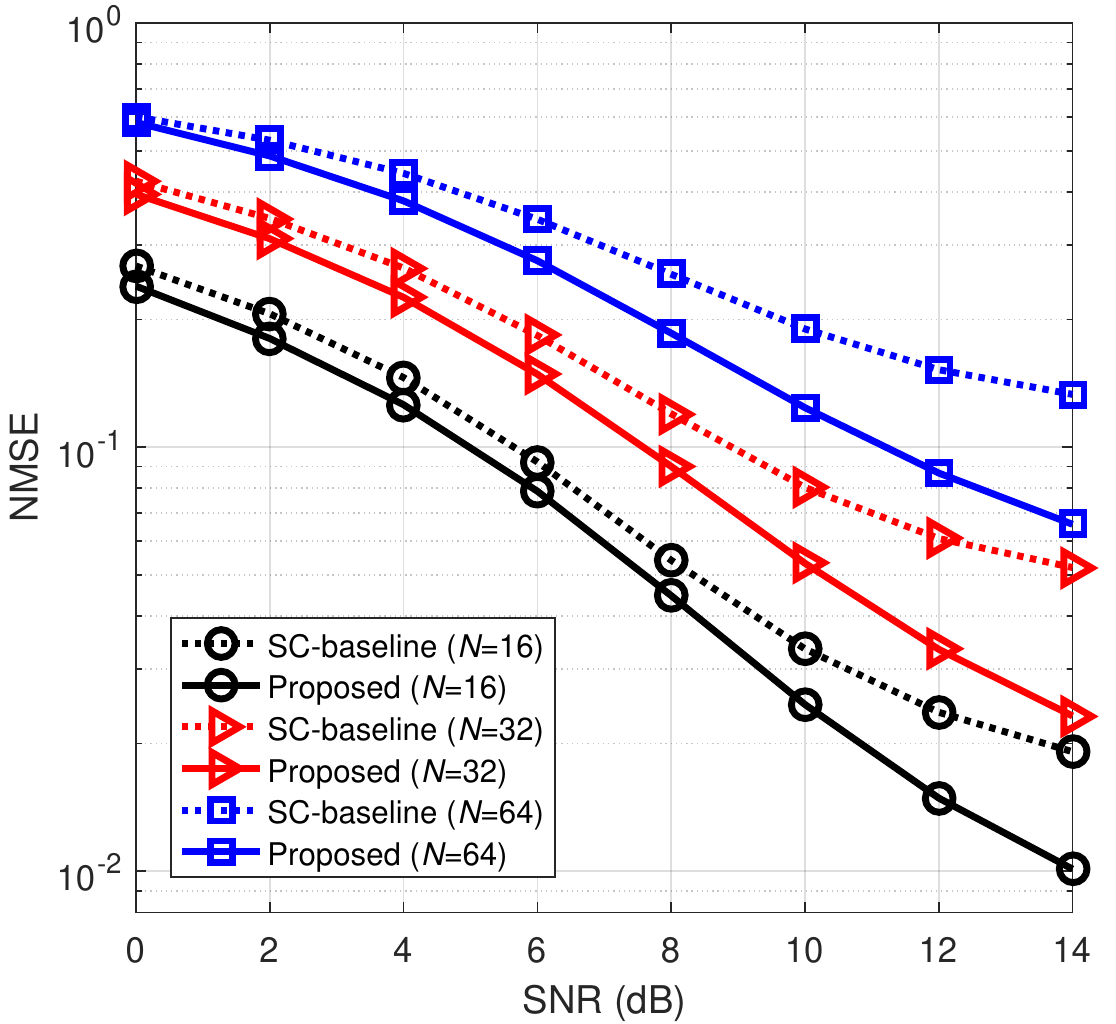}
\caption{ NMSE versus SNR, where $\rho = 0.2$, $M=512$.}
\label{fig3}
\end{figure}

In the NN training phase, the PPC $\rho$ and frame length (or UL-US length) $M$ are set to $\rho=0.2$ and $M=512$, respectively. Training set $\left\{ {{{{\mathbf{\tilde X}}}_u}} \right\}$ has 200,000 samples, and the batch size is 200 samples. During training, the SNR is set to $5$dB. We use Adam Optimizer as the training optimization algorithm \cite{b38} with parameters $\beta_1 = 0.99$ and $\beta_2 = 0.999$ \cite{b39}. The learning rates is set to 0.0001. The maximum number of iterations is 15,000. For each subnet training, the $L^2$ regularization \cite{b40} is adopted (see subsection 7.2.1 in  \cite{b40}). Three downlink CSI lengths (i.e., $N=16$, $N=32$, and $N=64$) are considered. Thus, three trained network models are obtained after training.

The testing data are generated by utilizing the same method of generating the training data. For $SNR\leq 10$dB, 200,000 testing samples are employed, while for $SNR > 10$dB, we stop the testing when at least 1000-bit errors are observed. For the SC-baseline method, three iterations are employed.

The training and testing of proposed method are carried out on a server with NVIDIA TITAN RTX GPU and Intel Xeon(R) E5-2620 CPU 2.1GHz$\times$16, and the results of SC-baseline are obtained by using Matlab simulation on the server CPU due to the lack of a GPU solution. With subnet-by-subnet training, each subnet in a network model (e.g., the model of $N=64$) is converged after 10,000 iterations. Totally, it takes no more than 80 minutes to train a whole network model (including four subnets), which is significantly faster than the case in \cite{b24} (about 3 days).

To verify the effectiveness of trained NN for the case where the test PPC and frame length are the same as that of training phase (i.e., $\rho=0.2$ and $M=512$), we first test the NMSE and BER performance and compare them against the SC-baseline. The performance curves are given in Fig. 3 and Fig. 4, respectively.

\begin{figure}
\centering
\includegraphics[scale=0.75]{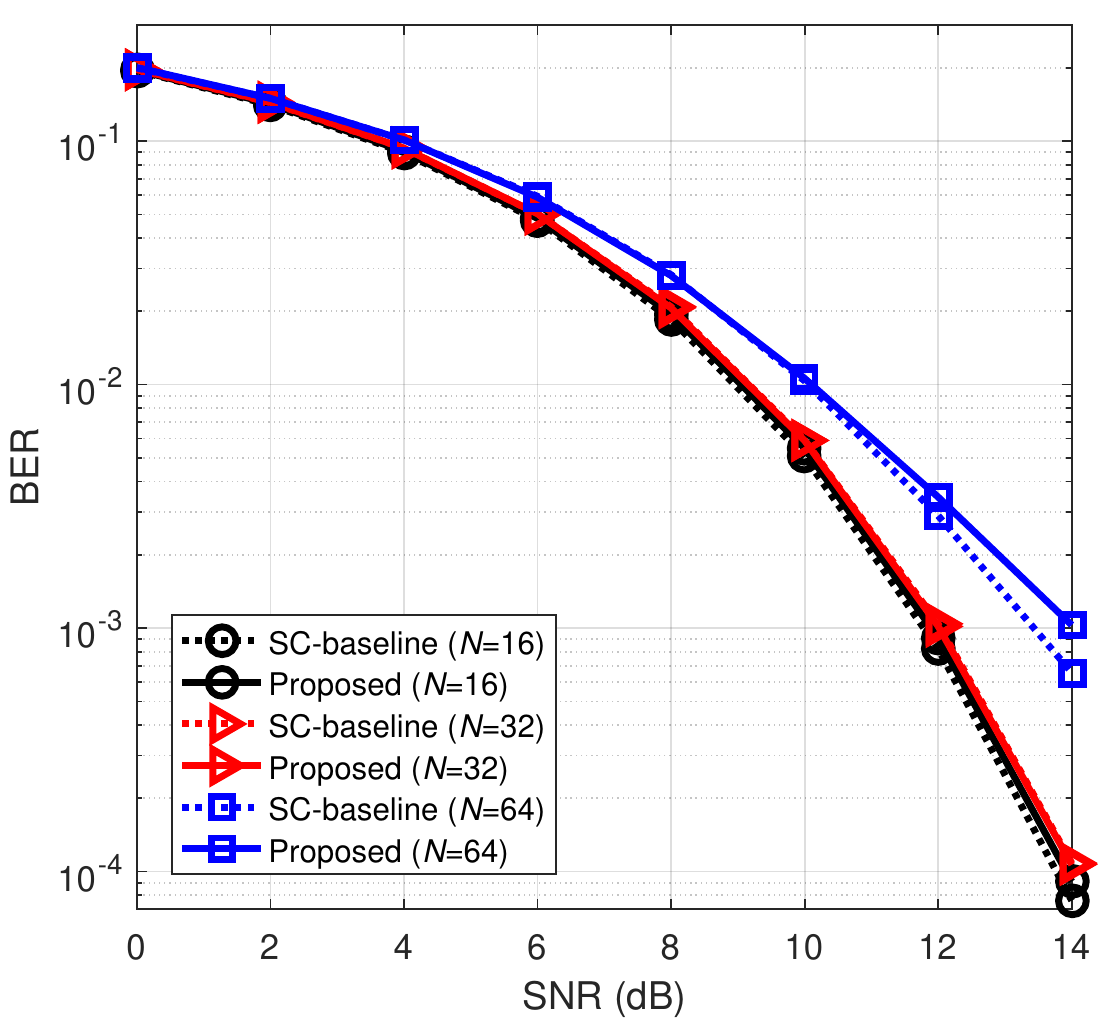}
\caption{BER versus SNR, where $\rho=0.2$, $M=512$.}
\label{fig4}
\end{figure}

Fig. 3 shows that the NMSE of each model (i.e., $N=16$, $N=32$, and $N=64$) outperforms the SC-baseline, especially at high SNR. Although $SNR=5$dB is adopted in training phase, the three trained network models work well in the entire SNR span varying from $0$dB to $14$dB. Thus, it is obvious that the designed and trained subnets (i.e., CSI-NET1 and CSI-NET2) have a good generalization ability for NMSE improvement.

In Fig. 4, the trained NNs and SC-baseline obtain almost identical BER when SNR is not greater than $10$dB. For the case where $N=64$ and $SNR \geq 12$dB, the BER of SC-baseline is slightly better than our trained NN. One reason for this is that a bigger $N$ would result in a smaller spreading gain and then deteriorate NN's learning ability. Another reason is likely that the testing SNR (14dB) is far from the training SNR (5dB). This can be confirmed that without changing the testing process, the NN trained at $SNR = 14$dB obtains similar testing BER as that of SC-baseline at 14dB. To resolve this kind of generalization degradation, the method that obtains training data from multiple SNRs in \cite{b24} can be used. Although the similar BER cannot be obtained when $N=64$ and $SNR \geq 12$dB, its BER performance in Fig. 4 is only slightly degraded. Especially, only one SNR (i.e., $SNR = 5$dB) is employed in our NN training, which bring us great benefits of practicality to avoid the difficulty of capturing multi-SNR data.

\begin{figure}
\centering
\includegraphics[scale=0.75]{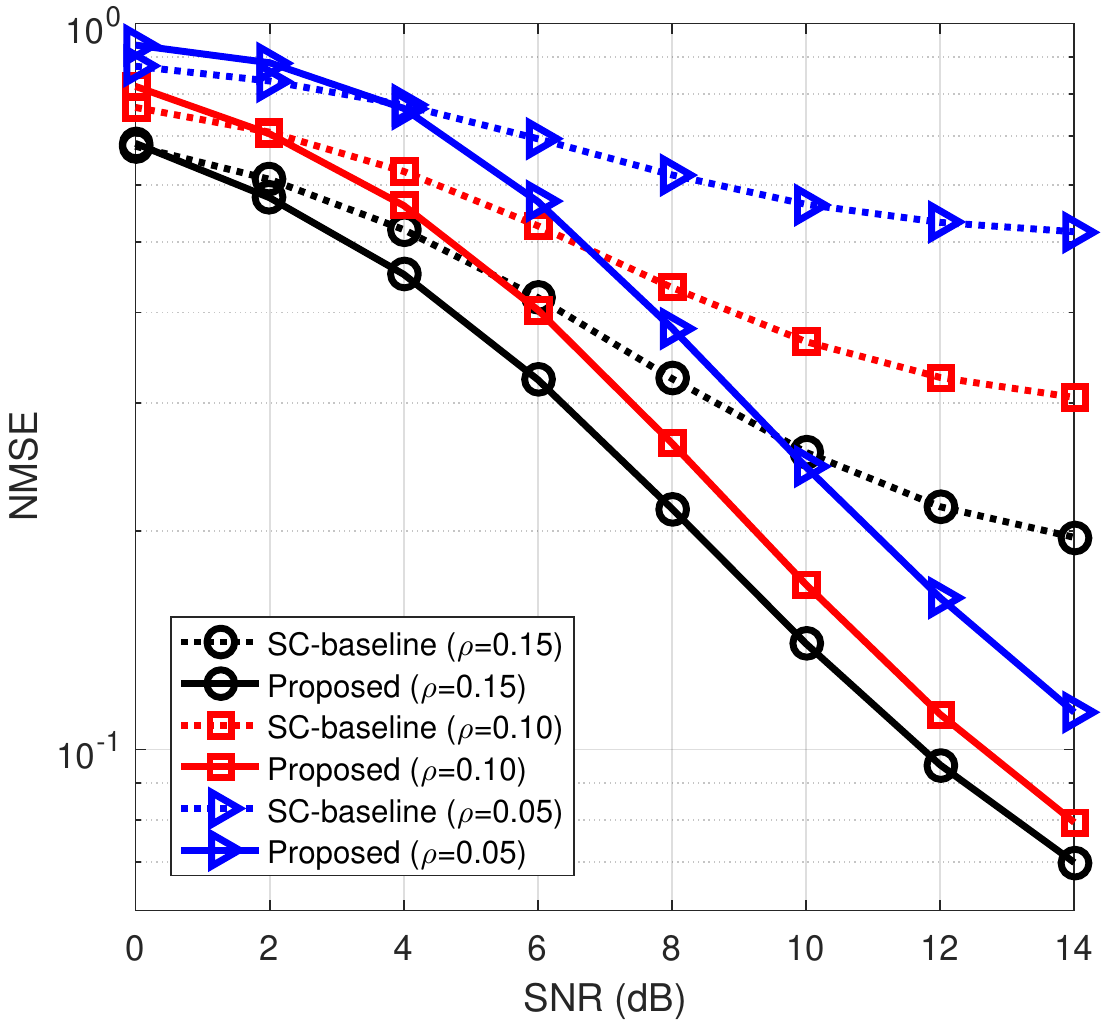}
\caption{NMSE versus SNR, where $N=64$, $M=512$.}
\label{fig5}
\end{figure}

\begin{figure}
\centering
\includegraphics[scale=0.75]{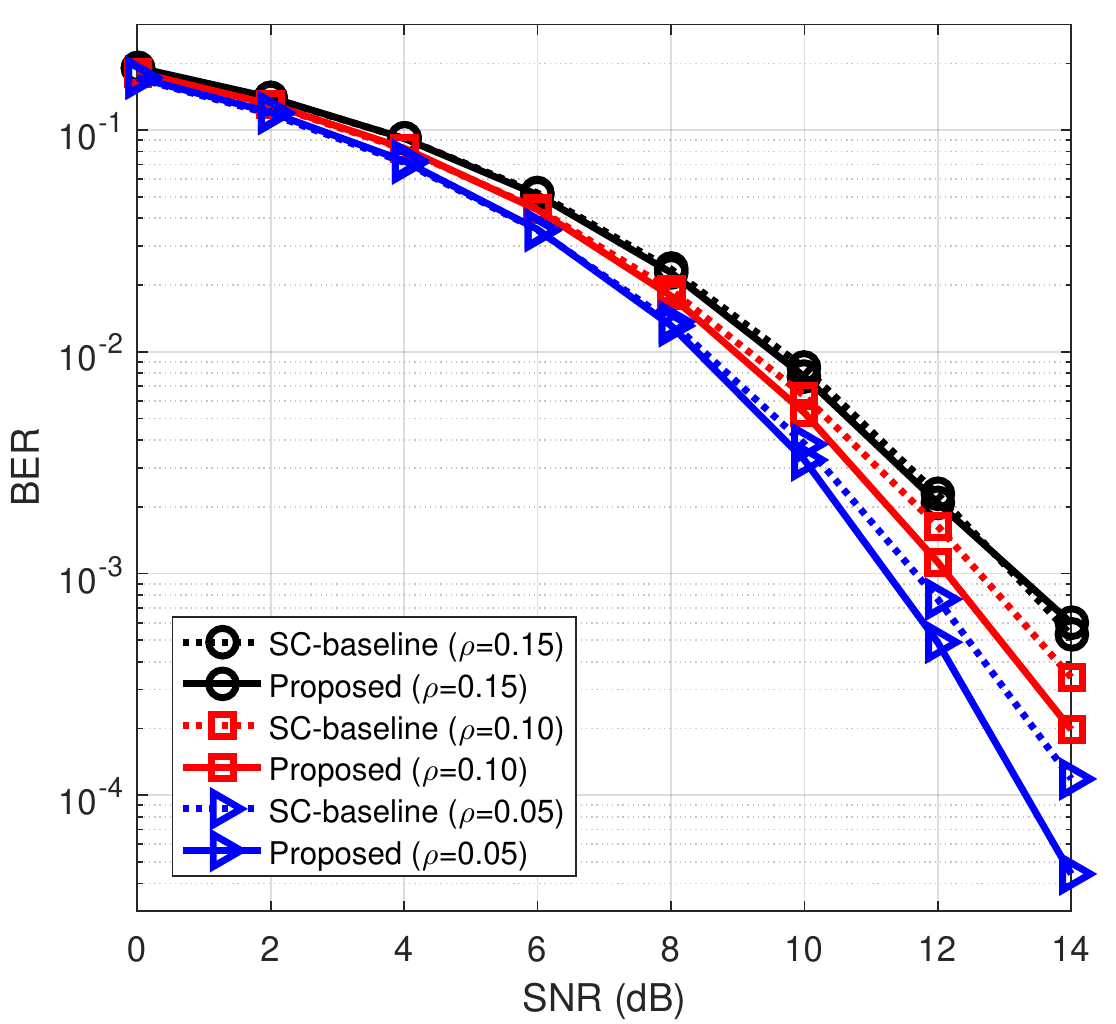}
\caption{BER versus SNR, where $N=64$, $M=512$.}
\label{fig6}
\end{figure}

\begin{figure}
\centering
\includegraphics[scale=0.75]{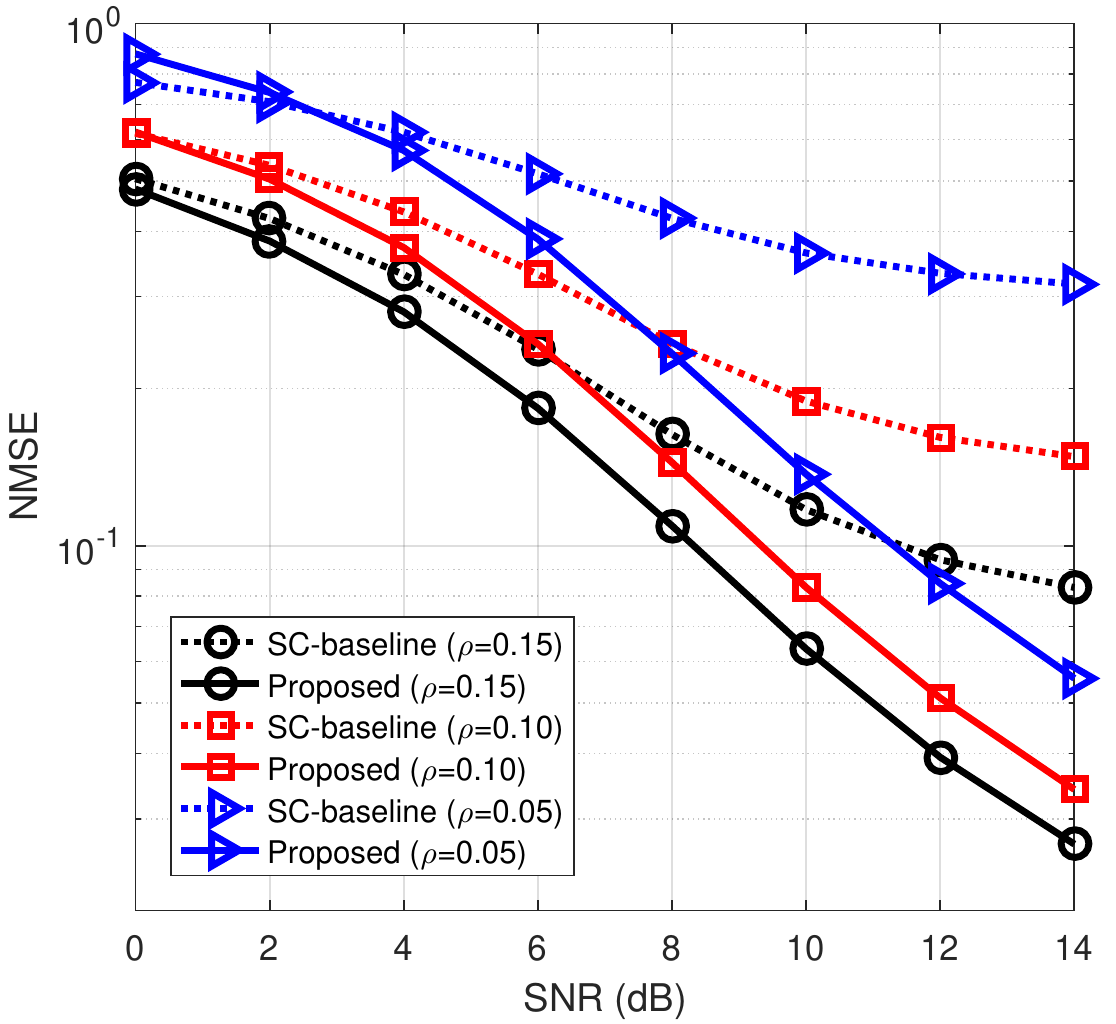}
\caption{NMSE versus SNR, where $N=32$, $M=512$.}
\label{fig7}
\end{figure}
\begin{figure}
\centering
\includegraphics[scale=0.75]{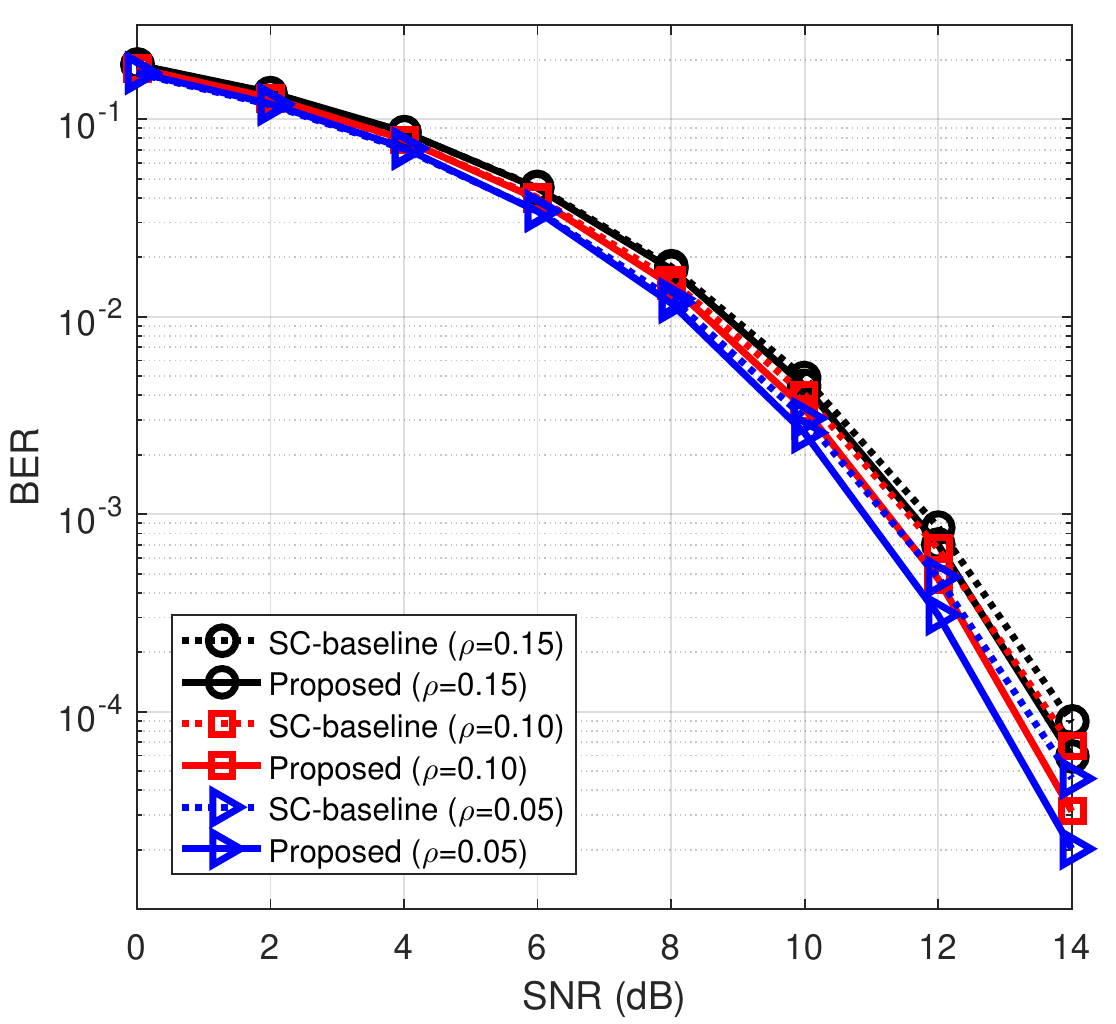}
\caption{BER versus SNR, where $N=32$, $M=512$.}
\label{fig8}
\end{figure}

To demonstrate the impact of PPC $\rho$ on the trained NNs, the BER and NMSE performances are given from Fig. 5 to Fig. 10. Note that, from Fig. 5 to Fig. 10, the NN training adopts $\rho=0.2$, while $\rho=0.05$, $\rho=0.10$, and $\rho=0.15$ are employed for testing. We use these simulations to illuminate that our NN has excellent generalization and robustness against the impact of PPC.

Given downlink CSI lengths $N=64$, 32, and 16, Fig. 5, Fig. 7 and Fig. 9 illustrate the NMSE performance with SNR varying from $0$dB to 14dB. Especially for relatively high SNR, e.g. $SNR\geq4$dB, it is obvious that the trained NNs evidently improve the NMSE when compared to SC-baseline. At the low SNR regime (e.g., $SNR \leq 2$dB) in Fig. 5 and Fig. 7, however, the NMSE of trained NNs is slightly inferior than that of SC-baseline. For example, in Fig. 7, the NMSE curve of the proposed method is a little higher than the baseline curve when $\rho = 0.05$ and $SNR \leq 2$dB. This situation is similar to that in Fig.~\ref{fig4}, where the decrease of spreading gain is a cause of the degradation of NN's learning ability. Although slightly inferior to the SC-baseline in certain low SNR regimes, our NN still shows prominent improvement in majority SNR regimes. On account of the training requirements (only one training PPC and one training SNR) and noise knowledge (without the knowledge of noise variance), the DL-based CSI feedback is still attractive.

To validate the generalization and robustness of BER against the impact of PPC, the BER performance is given in Fig. 6, Fig. 8 and Fig. 10 with $N=64$, $N=32$, and $N=16$, respectively. These figures reflect that, compared with the SC-baseline, our trained NN could achieve a similar or better BER performance. Especially, at the high SNR regime (e.g., $SNR \geq 10$dB), Fig. 6 shows BER improvement for the cases where $\rho = 0.05$ and $\rho = 0.10$ . A slight BER improvement is also observed in Fig. 8. The reason is likely that a small PPC avoids the generalization deterioration of BER performance due to the small superimposed interference from downlink CSI. It is worth noting that, the training PPC and SNR are fixed as $\rho = 0.2$ and $SNR = 5$dB, while the testing PPC and SNR are varying, e.g., $\rho = 0.05$, $0.10$ or $0.15$, and SNR is varying from $0$dB to $14$dB.

To sum up, compared to the SC-baseline, Fig. 3 to Fig. 10 show that the designed and trained multi-task network can improve the NMSE performance while keeping comparable (or better) BER performance. From Fig. 9 and Fig. 10, we can see that with similar BER, our NN can improve the NMSE for the case where $N=16$. As $N$ increase, it is observed from Fig. 5 and Fig. 6 (or Fig. 7 and Fig. 8) that, when $N=64$ (or $N=32$), both BER and NMSE of baseline can be improved, and a smaller PPC obtains greater improvements. Since we train three models under the conditions that $SNR = 5$dB, $\rho = 0.2$ and $M=512$, the designed NNs have a strong generalization ability for different SNRs and PPCs. In addition, the trained NN dose not need any knowledge of noise variance, which is also superior to the SC-baseline.

\section{CONCLUSIONS}
The accuracy of downlink CSI is the prerequisite of system capacity and link robustness. In this work, a CSI feedback method combined with SC and DL approaches is developed to improve the estimation of CSI in 5G wireless communication system without occupation of uplink bandwidth resource. We propose a multi-task neural network with subnet-by-subnet training method to facilitate the parameter tuning and expedite the convergence rate. The effectiveness of the proposed technique is confirmed by simulation result showing comparable or better NMSE and BER than that of baseline. This performance of the trained NN is also robust to varying SNR and PPC.

\begin{figure}
\centering
\includegraphics[scale=0.75]{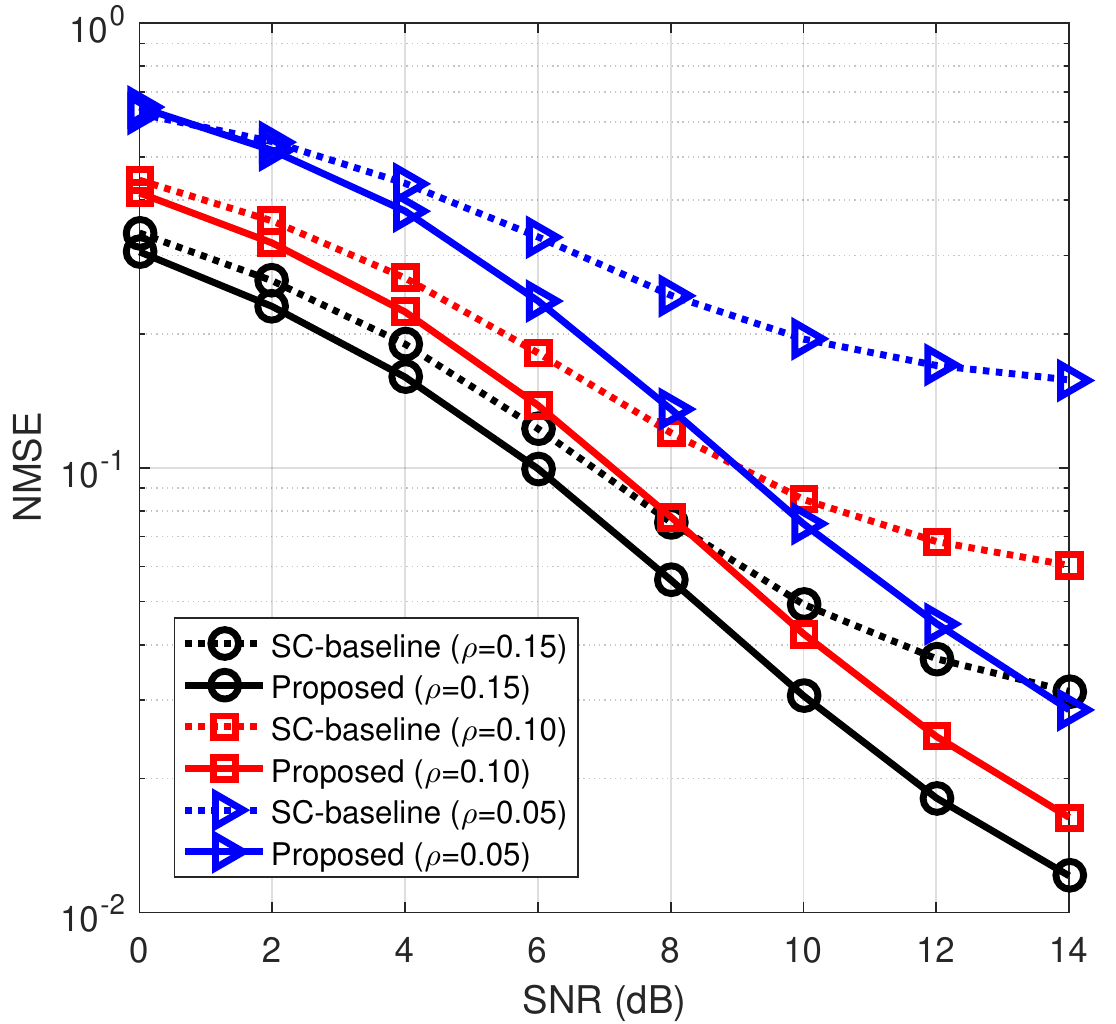}
\caption{NMSE versus SNR, where $N=16$, $M=512$.}
\label{fig9}
\end{figure}

\begin{figure}
\centering
\includegraphics[scale=0.76]{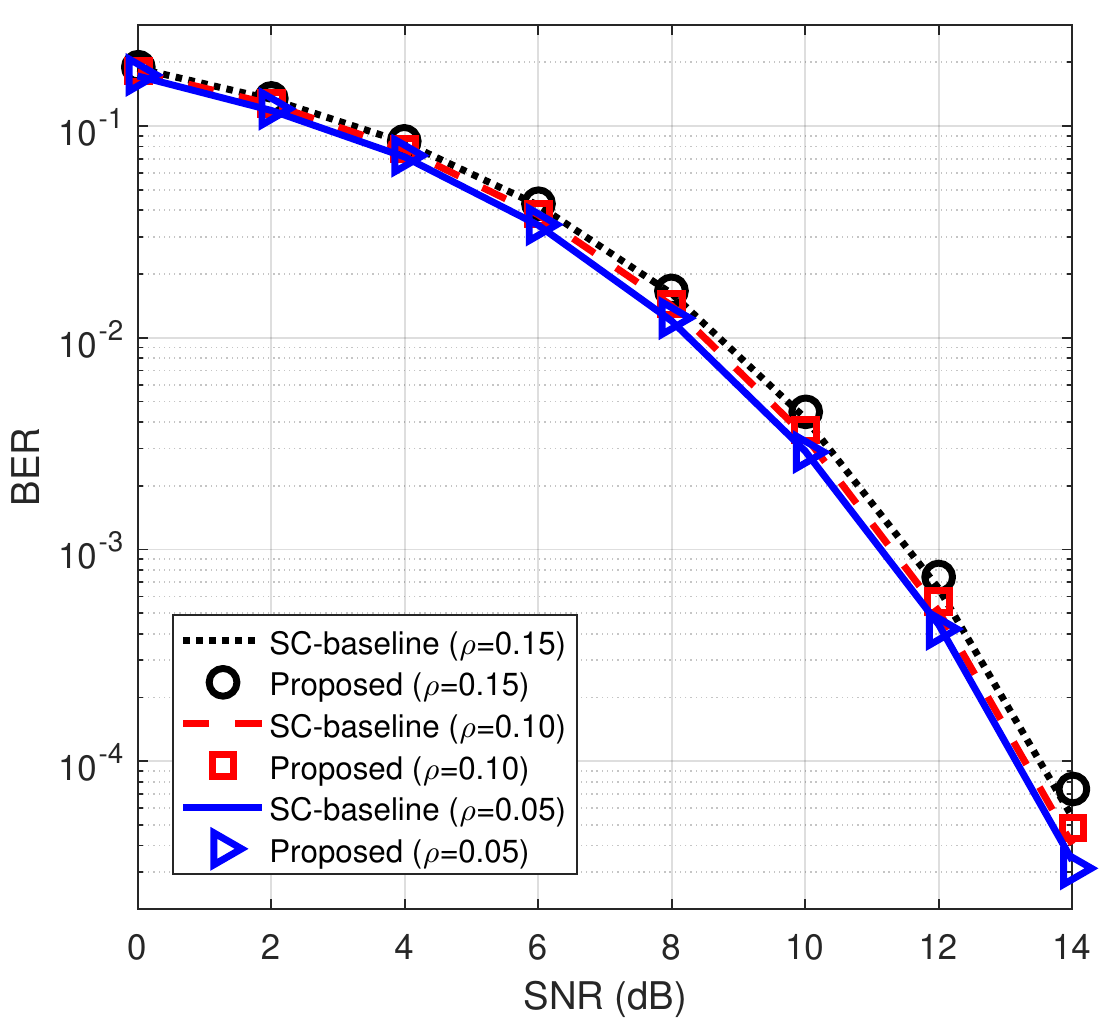}
\caption{BER versus SNR, where $N=16$, $M=512$.}
\label{fig10}
\end{figure}


\begin{IEEEbiography}[{\includegraphics[width=1in,height=1.25in,clip,keepaspectratio]{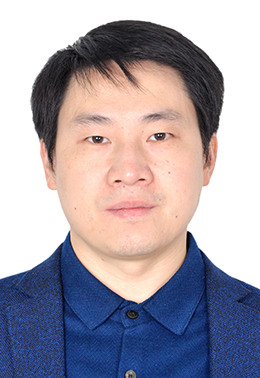}}]{Chaojin Qing} (M'15) received the B.S. degree in communication engineering from Chengdu University of Information Technology, Chengdu, China, in 2001, the M.S. and Ph.D. degrees in communications and information systems from the University of Electronic Science and Technology of China, Chengdu, China, in 2006 and 2011, respectively. From November 2015 to December 2016, he was a Visiting Scholar with Broadband Communication Research Group (BBCR) of the University of Waterloo, Waterloo, ON, Canada.

From 2001 to 2004, he was a teacher with the Communications Engineering Teaching and Research Office, Chengdu University of Information Technology, Chengdu, China. Since 2011, he has been an Assistant Professor with the School of Electrical Engineering and Electronic Information, Xihua University, Chengdu, China. He is the author of more than 40 papers and more than 20 chinese inventions. His research interests include detection and estimation, compressed sensing, and deep learning in physical layer of wireless communications.

\end{IEEEbiography}

\begin{IEEEbiography}[{\includegraphics[width=1in,height=1.25in,clip,keepaspectratio]{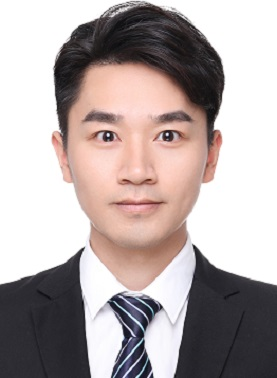}}]{Bin Cai}
received the B. S. degree from the School of Electrical Engineering and Electronic Information, Xihua University, Chengdu, China, in 2018, where he is currently  pursuing the M. S. degree under the supervision of Prof. Qing. His research interests include detection and estimation, channel state information feedback, and deep learning in physical layer of wireless communications.
\end{IEEEbiography}

\begin{IEEEbiography}[{\includegraphics[width=1in,height=1.25in,clip,keepaspectratio]{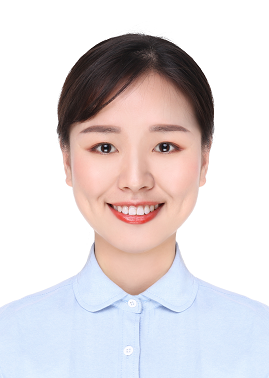}}]{Qingyao Yang} received the B. S. degree from the School of Electrical Engineering and Electrical Information, Xihua University, Chengdu, China, in 2017, where she is currently pursuing the M. S. degree under the supervision of Prof. Qing. Her research interests include compressed sensing, channel state information feedback, signal processing in wireless communications, and deep learning in physical layer of wireless communications.
\end{IEEEbiography}

\begin{IEEEbiography}[{\includegraphics[width=1in,height=1.25in,clip,keepaspectratio]{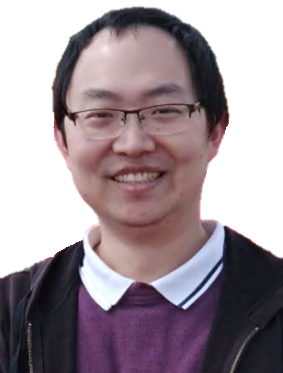}}]{Jiafan Wang} (S'15) received his B.S. degree and M.S. degree in Electrical Engineering from University of Electronic Science and Technology of China in 2006 and 2009, respectively. He accomplished the Ph.D.  degree in Computer Engineering at Texas A$\&$M University, College Station, TX, USA in 2017.

His major field of study is smart integrated circuit design, which includes multi-dimensional non-deterministic gate implementation with systematic optimization framework, self-training Analog/Digital Mixed-System for circuit feature calibration, and configurable locking mechanism against Analog IP piracy. He is currently working in Synopsys Inc. to develop the world's leading silicon chip design software in electronic design automation (EDA) industry.
\end{IEEEbiography}

\begin{IEEEbiography}[{\includegraphics[width=1in,height=1.25in,clip,keepaspectratio]{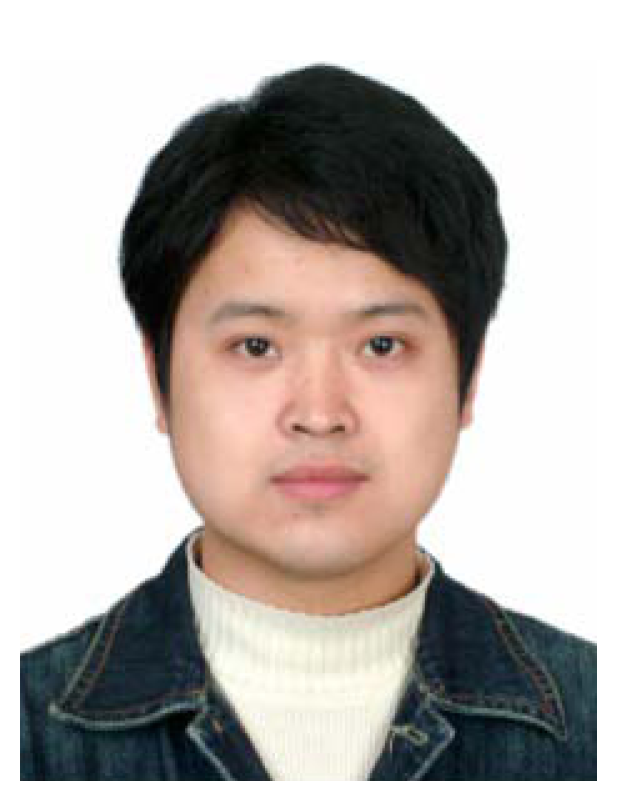}}]{Chuan Huang} (S'09--M'13) received the B.S. degree in math and the M.S. degree in communications engineering from the University of Electronic Science and Technology of China, Chengdu, and the Ph.D. degree in electrical engineering from Texas A$\&$M University, College Station, TX, USA, in 2012. From 2012 to 2013, he was with Arizona State University, Tempe, AZ, USA, as a Post-doctoral Research Fellow, and then promoted to Assistant Research Professor from 2013 to 2014. He was also a Visiting Scholar with the National University of Singapore and a Research Associate with Princeton University.

He is currently with the National Key Laboratory of Science and Technology on Communications, University of Electronic Science and Technology of China. His current research interests include energy harvesting communications, multicast traffic scheduling, full-duplex communications, and signal processing in wireless communications. He has served as a TPC member for many IEEE conferences.
\end{IEEEbiography}

\EOD

\end{document}